\documentstyle[preprint,prl,aps]{revtex}

\begin{document}
\title{Evidence for two distinct energy scales in the Raman spectra of YBa$_{2}$(Cu$%
_{1-x}$Ni$_{x}$)$_{3}$O$_{6.95}$}
\author{Y. Gallais$^{1}$, A. Sacuto$^{1}$, P. Bourges$^{2}$, Y. Sidis$^{2}$, A.
Forget$^{3}$ and D. Colson$^{3}$}
\address{$^{1}$Laboratoire de Physique du Solide (UPR 5 CNRS) ESPCI, 10 rue Vauquelin%
\\
75231 Paris, France\\
$^{2}$Laboratoire L\'{e}on Brillouin, CEA-CNRS, CE-Saclay, 91191\\
Gif-sur-Yvette, France\\
$^{3}$Service de Physique de l'Etat Condens\'{e}e, CEA-Saclay, 91191\\
Gif-sur-Yvette, France}
\date{\today}
\maketitle

\begin{abstract}
We report electronic Raman scattering from Ni-substituted $%
YBa_{2}Cu_{3}O_{6.95}$ single crystals with $T_{c}$ ranging from $\ 92.5$ K
to $78$ K. The fully symmetrical $A_{1g}$ channel and the $B_{1g}$ channel
which is sensitive to the $d_{x^{2}-y^{2}}$ gap maximum have been explored.
The energy of the $B_{1g}$ pair-breaking peak remains constant under Ni
doping while the energy of the $A_{1g}$ peak scales with $T_{c}$ ($%
E_{A_{1g}}/k_{B}T_{c}=5$). Our data show that the $A_{1g}$ peak tracks the
magnetic resonance peak observed in inelastic neutron scattering yielding a
key explanation to the long-standing problem of the origin of the $A_{1g}$
peak.
\end{abstract}

\noindent \qquad \newpage

\noindent \qquad The pairing mechanism that leads to high temperature
superconductivity in the cuprates is still a subject of intense controversy.
Electronic Raman Scattering (ERS) has been revealed as a very powerful tool
for probing selectively the electronic excitations in different regions of
the Fermi surface of the cuprates by orienting the incoming and outgoing
light polarizations\cite{Dev}. In particular ERS results in the
superconducting state of nearly stoichiometric cuprates advocate for a $%
d_{x^{2}-y^{2}}$ superconducting gap with its nodes along the $k_{x}=\pm \
k_{y}$ directions and its maxima along the $k_{x}=0$ and $k_{y}=0$
directions as a four-leaf clover \cite{Kang}\cite{Sacuto1}\cite{Staufer}. In
the $B_{1g}$ channel which probes the $k_{x}=0$ and $k_{y}=0$ directions
(where the magnitude of the gap $\Delta _{0}$ is maximum), the electronic
pair-breaking peak (corresponding to $2\Delta _{0}$) is observed in all
Raman spectra of optimally doped cuprates with energies between $8$ and $9$ $%
k_{B}T_{c}$ \cite{Staufer}\cite{Cooper}\cite{Chen}.\ However in the $A_{1g}$
channel which is sensitive to the entire Fermi surface, an unexpected,
strongly intense peak well below the $2\Delta _{0}$ energy is also observed
in all Raman spectra of cuprates near the optimally doped regime\cite
{Sacuto1}\cite{Staufer}\cite{Cooper}\cite{Chen}. An overview of the energy
locations of the $A_{1g}$ and $B_{1g}$ peaks detected by ERS measurements in
several cuprates with different $T_{c}$ is reported in Table 1. This table
reveals two distinct energy scales in the Raman spectra of cuprates. The $%
A_{1g}$ peak is located between $5$ and $6$ $k_{B}T_{c}$ and its strong
intensity challenges surprisingly the Coulomb screening present in the $%
A_{1g}$ channel. In fact, existing Raman theories based on the $%
d_{x^{2}-y^{2}}$ model fail to reproduce the intensity, the shape and the
position of the $A_{1g}$ peak\cite{Sacuto1}\cite{Wenger}. Expansion of the
Raman vertex to the second order of the Fermi surface harmonics has been
proposed to reproduce the relative $A_{1g}$ peak position with respect to
the $B_{1g}$ one\cite{Dev2}. However the major obstacle remains the Coulomb
screening which prevent us from reproducing the sharpness and the strong
intensity of the $A_{1g}$ peak.\ 

To go round this difficulty, strong mass fluctuations induced by several $%
CuO_{2}$ sheets has been suggested as well as a possible resonance effect of
the $A_{1g}$ Raman vertex which can enhance the $A_{1g}$ response\cite{Cardo}
\cite{Sherm}. These two last arguments are nevertheless hard to believe due
to the universal character of the $A_{1g}$ peak present in various optimally
doped cuprates even for cuprates with only one single $CuO_{2}$ sheet\cite
{Nemetschek}\cite{Kendzio}.\ Recent theoretical calculations suggest that
the $A_{1g}$ peak is controlled by a collective spin fluctuation mode which
is identified with the $Q=(\pm \pi ,\pm \pi )$ resonance observed by
Inelastic Neutron Scattering (INS) in the superconducting state\cite
{Venturini}. However no experimental proof has been provided up to now.

In this letter we report ERS experiments on Ni-substituted $YBCO$ systems
where Ni impurity is used as a probe for testing the physical origin of the $%
A_{1g}$ peak with respect to the $B_{1g}$ pair-breaking peak. Our most
striking result is that the $A_{1g}$ peak, contrary to the $B_{1g}$ one,
scales with $T_{c}$ with the same energy scale as the magnetic resonance
peak detected by INS\cite{Fong}\cite{Daï}\cite{Sidis}\cite{Bourges}. A
thorough examination shows that the temperature dependence of the $A_{1g}$
peak follows the same behavior as the INS resonance peak in contrast to the $%
B_{1g}$ peak. The question of a possible link between the $A_{1g}$ peak and
the INS peak will be addressed. In particular, the energy, the symmetry and
the non screening of the $A_{1g}$ peak are consistent with a spin
fluctuation origin for this peak.

ERS measurements were carried out on optimally doped Ni-substituted $%
YBa_{2}Cu_{3}O_{6.95}$ (YBCO) single crystals. They were grown in a gold
crucible using a flux technique\cite{Kaiser}. The Ni divalent ion is
believed to substitute preferentially for divalent copper in the $CuO_{2}$
plane of YBCO\cite{Hoffman}. Therefore Ni substitution offers a particularly
attractive way to reduce $T_{c}$ while preserving the carrier concentration 
\cite{Bobroff}. As a consequence Ni substitution in YBCO allows us to
investigate the $A_{1g}$ and $B_{1g}$ peaks as a function of $T_{c}$ near
the optimal doping where the $A_{1g}$ and $B_{1g}$ peaks are experimentally
the most easily observable. The $YBa_{2}(Cu_{1-x}Ni_{x})_{3}O_{6.95}$
twinned single crystals studied have Ni contents of $x=0,$ $x=0.01$ and $%
x=0.03$ with $T_{c}$ $=92.5$ K, $88$ K and $78$ K respectively.

Raman measurements were performed with a double monochromator using a single
channel detection and the $Ar^{+}$ and $Kr^{+}$ laser lines. The imaginary
parts of the Raman susceptibilities $\chi "(\omega )$ are deduced from the
raw spectra using the same procedure as in ref.\cite{Sacuto1}

The difference between the Raman responses obtained from $T=$ $130$ K
(normal state) and $T=35$ K (superconducting state) in the $B_{1g}$ and $%
A_{1g}$ channels as a function of Ni content are displayed in Fig. 1. The $%
\Delta \chi "=\chi "_{S}-\chi "_{N}$ \ spectra show for both $B_{1g}$ and $%
A_{1g}$ channels a set of sharp phonons peaks (typical width at half
maximum: $\sim 2$ meV) lying on a strong electronic background.\ In the $%
B_{1g}$ channel (Fig 1-a), for $x=0$, we observe in the $\Delta \chi "$
spectrum a strong and well-defined electronic peak at $67$ meV which is not
present in the normal state. On the other hand, the $\Delta \chi "$ spectrum
exhibits a negative intensity part below $40$ meV which corresponds to a
decrease of the electronic spectral weight in the superconducting state as
compared to the normal one. These two observations are the signature of the
opening of the superconducting gap along the $k_{x}=0$ and $k_{y}=0$
directions. The intensity of the $B_{1g}$ peak (defined as the value of $%
\Delta \chi "(\omega )$ at the maximum of the peak) decreases with Ni
content by a factor of $2$ for $x=0.01$ and $3$ for $x=0.03$. A striking
feature is that the $B_{1g}$ peak does not shift in energy with increasing
Ni content. Even though the $B_{1g}$ peak becomes broader (width at half
maximum increases by a factor of $1.5$) its position remains centered at $67$
meV which suggests that the gap maximum is unaffected despite the
introduction of $x=0.03$ Ni impurities and the $T_{c}$ decrease of $14$ K.

In the $A_{1g}$ channel (Fig 1-b), we clearly detect in the $\Delta \chi "$
spectra for $x=0$ a broad and intense peak in the continuum centered around $%
40$ meV. The $A_{1g}$ peak, like the $B_{1g}$ one, appears only in the
superconducting state. However, it is much more robust than the $B_{1g}$
peak and is still well defined even for the highest nickel concentration ($%
x=0.03$) with an intensity decrease less than a factor of $1.6$. In contrast
with the $B_{1g}$ peak which does not change its position, the $A_{1g}$ peak
shifts significantly to lower energy with increasing Ni content. It softens
from $40$ meV at $x=0$ down to $38$ meV and $33$ meV for $x=0.01$ and $%
x=0.03 $ respectively.

Fig. 2 displays the ERS $A_{1g}$ and $B_{1g}$ peaks energies as a function
of $T_{c}$. Our data clearly shows that while the $B_{1g}$ peak energy
remains constant upon Ni substitution (i.e. the magnitude of the gap at ($%
\pi ,0$) does not scale with $T_{c}$), the $A_{1g}$ peak energy scales with $%
T_{c}$. In fact if we compare our results with INS data obtained from
Ni-substituted optimally doped $YBCO$\cite{Sidis} we find quantitative
correspondence between the $A_{1g}$ peak and the INS magnetic resonance. The
ERS $A_{1g}$ peak and the INS resonance energies both scale with $T_{c}$ and
can be plotted along the same line corresponding to $E/k_{B}T_{c}=5$. This
similarity does not only hold for $Y-123$ system since a quite good
correspondence between the energies of the $A_{1g}$ peak and the INS
resonance is also achieved for the $Bi-2212$ and $Tl-2201$ systems as shown
in Table 1.

To go further, we have investigated the temperature dependence of the ERS $%
A_{1g}$ and $B_{1g}$ peaks with respect to their intensity and their energy
location in the pure optimally doped $YBCO$\ system ($x=0$) and performed a
comparative analysis with the INS resonance. Fig. 3-a shows that the ERS $%
B_{1g}$ peak intensity is linear as a function of temperature whereas the
ERS\ $A_{1g}$ peak and INS resonance \cite{Bourges}\ intensities decrease in
the form of a step function and vanish at $T_{c}$. In addition the energy
positions of the $A_{1g}$ peak and the INS resonance remain constant with
the same value under the temperature variation as shown in Fig. 3-b.

In Fig. 4 the $A_{1g}$ responses as a function of various excitations lines
are displayed. It appears that the $A_{1g}$ peak position is not sensitive
to the laser line excitations and therefore cannot be attributed to a
resonance effect of the $A_{1g}$ Raman vertex.

In the light of these new results we discuss the origin of the $A_{1g}$
peak. On one hand, our data have clearly shown that the ERS $A_{1g}$ and $%
B_{1g}$ peaks exhibit drastically different behaviors as a function of $%
T_{c} $ (for Ni- substituted $YBCO$ systems) but also as a function of
temperature (for pure $YBCO$ system). On the other hand we have clearly
established quantitative correspondences between the ERS $A_{1g}$ peak and
the INS magnetic resonance peak. These experimental results suggest that in
contrast to the attribution of the $B_{1g}$ peak as arising from the pair
breaking process in the charge channel, the $A_{1g}$ peak could be coming
from a different origin. The idea of a magnetic origin for the $A_{1g}$ peak
associated with the anti-ferromagnetic (AF) spin fluctuations is very
appealing for three main reasons. Firstly, because we have introduced the
experimental evidence that the $A_{1g}$ peak and the INS magnetic resonance
are deeply related. Secondly, the Coulomb screening is expected to be
inefficient because AF fluctuations are not coupled with the electron
density. This explains why the $A_{1g}$ peak detected in the Raman spectra
is unscreened. Finally, the resonance peak has the four-fold symmetry in the 
$k-$space which corresponds to a fully symmetrical $A_{1g}$ configuration.
It follows that the magnetic resonance contribution to the Raman response
will be relevant in the $A_{1g}$ channel only\cite{Venturini}.

In conclusion, through Ni-substitution we have clearly established the same
energy scale $5$ $k_{B}T_{c}$ for the INS resonance and the ERS $A_{1g}$
peak and we have shown that the ERS\ $A_{1g}$ and INS resonance peaks follow
the same temperature dependence. This $A_{1g}$ energy scale is different
from the one of the pair breaking process related to the $B_{1g}$ peak. The
existence of an additional energy scale in the superconducting state,
observed both in ERS and INS experiments can provide the key towards an
understanding of the pairing mechanism in high-$T_{c}$ superconducting
cuprates.

\bigskip

We thank P. Monod, E. Ya. Sherman, F. Venturini, T. P. Devereaux and F.
Rullier-Albenque for very fruitful discussions and S. Poissonnet for the
chemical anlysis by electron probe.

{\bf FIGURE\ CAPTIONS}

FIG. 1. Raman response functions difference between the superconducting and
normal states of $YBa_{2}(Cu_{1-x}Ni_{x})_{3}O_{6.95}$ for $x=0$, $0.01$ and 
$0.03$ in the $B_{1g}$ channel (a) and $A_{1g}$ channel (b). The $514.52$ nm
laser line was used. The peaks were fit to a gaussian shape in their upper
parts (bold lines) to define their positions (the mathematical continuation
is shown as a dotted line).

FIG. 2. Raman and neutron peak energies\cite{Sidis} as a function of the
critical temperature $T_{c}$ in $YBa_{2}(Cu_{1-x}Ni_{x})_{3}O_{6.95}$ for $%
x=0$, $0.01$ and $0.03$. The horizontal line for the $B_{1g}$ peak positions
is just a guide to the eye while the $A_{1g}$ peak and the neutron resonance
are fitted by a straight line representing $E/k_{B}T_{c}=5.$

FIG. 3. Temperature dependence of the intensities (a) and energy positions
(b) of the ERS $A_{1g}$, $B_{1g}$ peaks and the INS resonance\cite{Bourges}
in optimally doped $YBa_{2}Cu_{3}O_{7-\delta }$. The INS resonance
intensities (arbitrary units) have been normalized to the ERS peaks
intensities. The dotted lines are guide to the eye.

FIG.\ 4. Raman response functions in the $A_{1g}+B_{2g}$ channel obtained
from different excitation lines at $T=$ $35$ K in a pure optimally doped
YBCO.

{\bf TABLE\ CAPTION}

TABLE\ I. An overview of the peak locations of the ERS $A_{1g}$ ($E_{A_{1g}}$%
)and $B_{1g}$ ($E_{B_{1g}}$)and the INS resonance($E_{R}$). $a$ this study, $%
b$ ref.\cite{Fong} , $c$ ref. \cite{Misochko}, $d$ ref.\cite{Fong2} ($%
T_{c}=91$ K), $e$ ref.\cite{Chen}, $f$ ref.\cite{Gasparov}, $g$ ref.\cite{He}
($T_{c}=92$ K), $h$ ref.\cite{Sacuto1}. $\ast $ The neutron magnetic
resonance peak has never been reported in $LSCO$. However, the temperature
dependence of the neutron magnetic intensity observed for $x=0.14$ at the
incommensurate wave vector around $9-16$ meV \cite{Mason} exhibits a certain
similarity with that of the neutron resonance observed in the other high-$%
T_{c}$ cuprates.


\bigskip

\begin{references}
\bibitem{Dev}  T. P. Devereaux et al., Phys. Rev. Lett. {\bf 72}, 396 (1994).

\bibitem{Kang}  Moonsoo Kang, G. Blumberg, M. V. Klein and N. N. Kolesnikov,
Phys. Rev. Lett. {\bf 77}, 4434, (1996).

\bibitem{Sacuto1}  A.\ Sacuto, J. Cayssol, D.Colson and P.\ Monod, Phys.
Rev. B {\bf 61}, 7122 (2000).

\bibitem{Staufer}  T. Staufer et al., Phys. Rev. Lett. {\bf 68}, 1069 (1992).

\bibitem{Cooper}  S. L. Cooper et al., Phys. Rev. B {\bf 38}, 11934 (1988).

\bibitem{Chen}  X. K. Chen et al., Phys. Rev. Lett. {\bf 73}, 3290 (1994).

\bibitem{Wenger}  F. Wenger and M.\ K\"{a}ll, Phys. Rev. B {\bf 55}, 97
(1997).

\bibitem{Dev2}  T. P. Devereaux and D. Einzel, Phys. Rev. B {\bf 51}, 16336
(1995); {\bf 54}, 15547 (1996).

\bibitem{Cardo}  M. Cardona, T. Strohm and X. Zhou, Physica C {\bf 282-287},
222 (1997).

\bibitem{Sherm}  E. Ya. Sherman and C. Ambrosch-Draxl, Solid State Comm. 
{\bf 115}, 669 (2000).

\bibitem{Gasparov}  L. V. Gasparov et al. Phys. Rev. B {\bf 55}, 1223 (1997).

\bibitem{Nemetschek}  Nemetschek et al. Phys. Rev. B {\bf 47}, \ 3450 (1993).

\bibitem{Kendzio}  C. Kendziora et al. Physica C {\bf 341-348}, 2189, (2000).

\bibitem{Venturini}  F. Venturini, U. Michelucci, T. P. Devereaux and A. P.
Kampf, Phys. Rev. B {\bf 62}, 15204 (2000).

\bibitem{Fong}  H. F. Fong et al., Phys. Rev. B {\bf 61}, 14773 (2000).

\bibitem{Daï}  P. Da\"{i}, H. A. Mook, R. D.\ Hunt and F. Dogan, Phys. Rev.
B {\bf 63}, 054525 (2001).

\bibitem{Sidis}  Y. Sidis et al., Phys. Rev. Lett. {\bf 84}, 5900 (2000).

\bibitem{Bourges}  P. Bourges, L. P. Regnault, Y. Sidis and C. Vettier,
Phys. Rev. B {\bf 53}, 876 (1996).

\bibitem{Kaiser}  D. L. Kaiser, F. Holtzberg, B. A. Scott, T. R. McGuire,
Appl. Phys. Lett. {\bf 51},1040 (1987).

\bibitem{Hoffman}  S. A. Hoffman, M. A. Castro, G. C. Follis and S. M.
Durbin, Phys. Rev. B {\bf 49}, 12170 (1994).

\bibitem{Bobroff}  J. Bobroff et al., Phys. Rev. Lett. {\bf 79}, 2117 (1997).

\bibitem{Misochko}  O. V. Misochko and G. Gu, Phys. Rev. B {\bf 59}, 11183
(1999).

\bibitem{Fong2}  H. F. Fong et al., Nature {\bf 398}, 588 (1999).

\bibitem{He}  H. F. He et al. to be published.

\bibitem{Mason}  T. E. Mason et al., Phys. Rev. Lett. {\bf 77}, 1604 (1996).
\end{references}
\end{document}